\documentclass{article}

\usepackage{spconf,amsmath,graphicx}
\usepackage{caption}
\usepackage{subcaption}
\usepackage{multirow}
\usepackage{multicol}
\usepackage{graphicx}
\usepackage{microtype}
\usepackage{graphicx}
\usepackage{booktabs} 
\usepackage{amssymb}
\usepackage{hyperref}
\usepackage{threeparttable, multirow}

\newcommand{\inparanth}[1]{\left(#1\right)}
\newcommand{\mcalp}{\mathcal{P}}

\title{Flexible Keyword Spotting based on Homogeneous Audio-Text Embedding}
\name{Kumari Nishu, Minsik Cho, Paul Dixon, Devang Naik}
\address{Apple}
\begin{document}
%
\maketitle

\begin{abstract}

Spotting user-defined/flexible keywords represented in text frequently uses an expensive text encoder for joint analysis with an audio encoder in an embedding space, which can suffer from heterogeneous modality representation (i.e., large mismatch) and increased complexity.
In this work, we propose a novel architecture 
to efficiently detect arbitrary keywords based on an audio-compliant text encoder which inherently has homogeneous representation with audio embedding, and it is also much smaller than a compatible text encoder. 
Our text encoder converts the text to phonemes using a grapheme-to-phoneme (G2P) model, and then to an embedding using representative phoneme vectors, extracted from the paired audio encoder on rich speech datasets.
We further augment our method with confusable keyword generation to develop an audio-text embedding verifier with strong discriminative power.
Experimental results show that our scheme outperforms the state-of-the-art results on Libriphrase hard dataset, increasing Area Under the ROC Curve (AUC) metric from 84.21\% to 92.7\% and reducing Equal-Error-Rate (EER) metric from 23.36\% to 14.4\%.

\end{abstract}


\section{Introduction}
\label{sec:intro}


Keyword spotting (KWS) is the task of detecting intended keywords from spoken speech.
KWS can be  classified into fixed KWS, where only known keywords are targeted~\cite{fixed_kws1, fixed_kws2, fixed_kws3_Triplet_Loss}, and more challenging user-defined/flexible KWS where arbitrary textual keywords need to be detected~\cite{flexible_kws2_asr_embedding, Shin2022LearningAA, nishu2023matching}.
To spot arbitrary keywords  accurately, prior arts either require enrolling the keywords as speech signals  or rely on the joint analysis from audio and text encoders.

\begin{figure}[t!]
  \centering
        \begin{subfigure}[b]{0.45\linewidth}
         \centering
         \includegraphics[width=\linewidth]{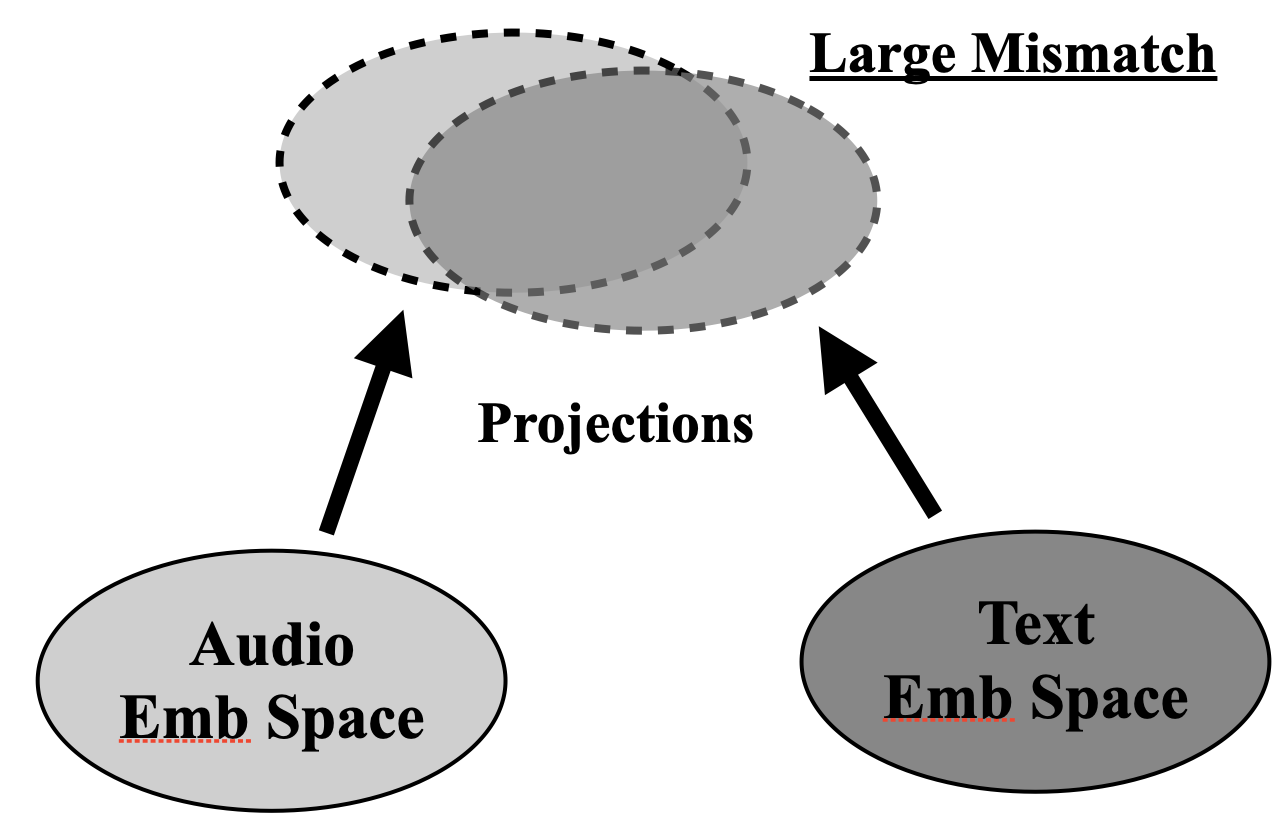}
         \caption{A projection method can create the large mismatch between audio and text embedding spaces~\cite{Shin2022LearningAA, nishu2023matching}.}
         \label{fig:large_mismatch}
     \end{subfigure}
     \hspace{0.1 in}
     \begin{subfigure}[b]{0.45\linewidth}
         \centering
         \includegraphics[width=\linewidth]{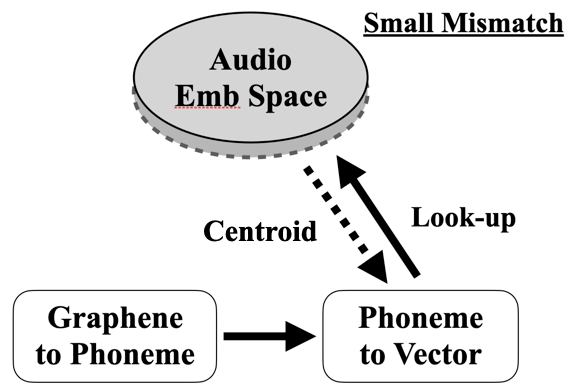}
         \caption{Phoneme-to-Vector can reduce the mismatch between the audio and text embedding spaces.}
         \label{fig:small_mismatch}
     \end{subfigure}
 \caption{Our proposed technique generates  text embedding of user-defined keywords using the phoneme-to-vector which is also built with the paired audio encoder, making the mismatch between two embedding spaces small and improving the performance of flexible KWS, without an extra text encoder.}
 
 \label{fig:mismatch}    
 \vspace{-0.2in}
\end{figure}

\begin{figure*}[t!]
  \centering
  \includegraphics[width=\textwidth]{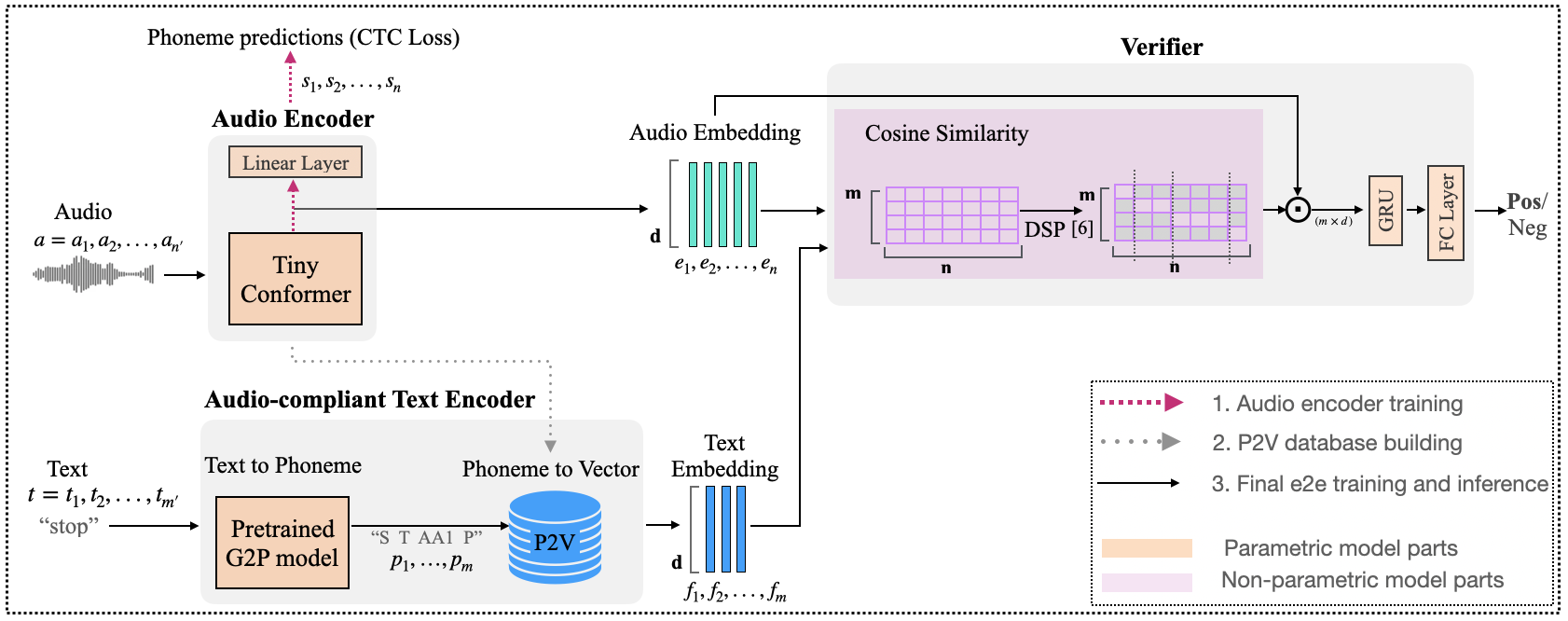}
  \caption{Overall architecture of our proposed model CED (Common Embedding based Detector). Input is an audio-text pair and output is a verification score indicating if the audio content is same as the text.}
  \label{fig:model}
\end{figure*}

Prior works \cite{flexible_kws2_asr_embedding, fixed_kws3_Triplet_Loss, Jung2022MetricLF} have used enrolled audio samples for user-defined KWS, broadly refereed as query-by-example methods. \cite{flexible_kws2_asr_embedding} relies on ASR model for the embeddings of enrolled and query audio. In~\cite{dtw, dtw_embedding, dtw_embedding1}, Dynamic Time Warping (DTW) was used to measure the similarity between the enrolled and the query embedding. On the other hand, \cite{Shin2022LearningAA, nishu2023matching} accept the keyword enrollment in text which offers a better user interface and most compatible practical applications with streaming scenarios for low-powered keyword detection.
 

Although the text encoder based schemes  \cite{Shin2022LearningAA, nishu2023matching} may avoid a cumbersome enrollment process~\cite{flexible_kws2_asr_embedding, fixed_kws3_Triplet_Loss, Jung2022MetricLF}, they face challenges. These include the possibility of significant embedding mismatch due to the use of two encoders representing heterogeneous modalities, which can lead to reduced accuracy. Moreover, jointly processing embedding vectors from independently trained modality encoders (thus in different spaces) requires a transform mechanism, like projection, to bring the audio and text embeddings to a joint space, leading to increased parameters and a larger package size during deployment.


To address such challenges, we propose a novel flexible KWS powered by an audio-compliant text encoder. Our text encoder is based on a grapheme-to-phoneme (G2P) model ~\cite{g2pE2019} and a phoneme-to-vector (P2V)  table which is constructed with the most general representation of each phoneme from the paired audio encoder. The text embedding is  synthesized by efficiently concatenating the phoneme vectors from P2V for the word analyzed by the G2P.
Hence, our text encoder inherently produces the text embedding in the same space as the audio encoder, thus eliminating extra transformation and underlying mismatch problems.
Along with confusable keyword generation, our technique showed state-of-the-art results on flexible KWS. 
Fig.~\ref{fig:mismatch} depicts the novelty of our work over the prior arts, and 
our  contributions are  following:

\begin{itemize}
\vspace{-0.05in}\item  We propose a   non-parametric audio-compliant text encoder, to produce text embedding derived from the learnt phonetic embedding space of the audio encoder. 
\vspace{-0.05in}\item We introduce a   confusable keyword generation scheme  to reduce false triggers by making flexible KWS aware of  real-word phonetic confusability.
\vspace{-0.05in}\item We formulate a discriminative setting to train an end-to-end audio-text based KWS model consisting of an audio encoder, an audio-compliant text encoder, and a verifier.
\end{itemize}

\section{Proposed Method}
\label{sec:proposed_method}

In this section, we describe our proposed methodology, Common Embedding based Detector (CED), shown in Fig. \ref{fig:model}. CED consists of three broader modules - \textit{audio encoder}, \textit{audio-compliant text encoder}, and a \textit{verifier}. The CED model takes an audio and keyword text pair as input and checks if they are a match. The CED model is trained in three steps - audio encoder training, Phoneme-to-Vector (P2V) database building, and final end-to-end training of CED. 

We denote an input sample as $\inparanth{a,t,l}$, where audio 
$a=\inparanth{a_1, a_2, \dots, a_{n'}}$ is a sequence of audio frames, text $t=\inparanth{t_1, t_2, \dots, t_{m'}}$ is a sequence of words, and $l$ is the binary label where $l=1$ represents a positive input pair.

\subsection{Audio Encoder}
\label{subsec:Audio_Encoder}
We use a small conformer \cite{Gulati2020ConformerCT} architecture as an audio encoder which bundles self-attention \cite{Vaswani2017AttentionIA} layers and convolutional layers, capturing both global and local audio contexts. We train the conformer for the phoneme prediction task which allows us to build homogeneous audio-text encoders, and create text embedding in the same phonetic embedding space. This mitigates the mismatch issue as compared to using a generic text encoder \cite{nishu2023matching}.




In detail, we train the audio encoder using CTC loss \cite{Graves2006ConnectionistTC} at the first step, as shown in Fig. \ref{fig:model}. Then, we use the trained audio encoder to build the Phoneme-to-Vector (P2V) database detailed in Section \ref{subsec:Text_Encoder}. Finally,
we perform the end-to-end discriminative training of CED model for the KWS task.
For an input audio $a$, Let us denote the embedding output from the audio encoder as $e=\inparanth{e_1, e_2, \dots, e_n}$, where $e_i\in R^{d}, \forall\, i=1,2,\dots, n$ and $n < n'$ as the conformer module includes a subsampling layer which reduces the input sequence length.


\subsection{Audio-compliant Text Encoder}
\label{subsec:Text_Encoder}
In this section, we describe our novel \textit{audio-compliant text encoder}
whose purpose is to replace a generic text encoder with our light-weight and embedding-sharing encoder to deliver high-quality flexible KWS. The key idea is to derive the text embedding from the learnt phonetic embedding space of the paired audio encoder with a much smaller model footprint.


\textbf{Text to Phoneme:} The user-defined keywords are enrolled as text, shown in Fig. \ref{fig:model}. To handle audio and text both in the same phonetic embedding space, we first need to convert the text graphemes (spelling) to phonemes (pronunciation) sequence. We use a pre-trained G2P (grapheme-to-phoneme) \cite{g2pE2019} model for this conversion. Let $\mcalp$ denotes the set of $74$ phonemes from the G2P model vocabulary. For an input text $t$, 
the G2P model generates the output $p=\inparanth{p_1, p_2, \dots, p_m}$ (where $p_i\in \mcalp, \forall\, i \in \{1,2,\dots, m\}$), which will be  directed to the Phoneme-to-Vector database to synthesize text embedding.


\textbf{Phoneme to Vector:} Phoneme-to-Vector (P2V) converts given phonemes into vectors which are further concatenated to yield a text embedding, and such conversion is based on P2V database as shown in Fig. \ref{fig:model}. Hence, a good P2V database is crucial for performant KWS.
To build the P2V database, we run the trained audio encoder on the Libriphrase training dataset \cite{Shin2022LearningAA} in evaluation mode (see Section~\ref{sec:experiments} for details).

\textbf{Sampling:} For a sample input $\inparanth {a, t}$, the conformer block in Fig. \ref{fig:model} produces an audio embedding   $e$ which is passed to the last linear layer of the audio encoder to produce phoneme prediction scores across all phonemes for each audio frame. 
We denote this score as $s=\inparanth{s_1, s_2, \dots, s_n}$, where $s_i\in \left[0,1\right]^{|\mcalp|}, \forall\, i \in \{1,2,\dots, n\}$. Using $s$, we perform a greedy decoding to get the predicted phoneme sequence: by simply taking the maximum probable phoneme at each audio frame and removing consecutive duplicate phonemes. The predicted phoneme sequence becomes $\hat p = \inparanth{\hat p_{j_1}, \hat p_{j_2}, \dots, \hat p_{j_{m'}}}$, where $1=j_1<j_2<\dots<j_{m'}\leq n$. 
We measure the quality of $\hat{p}$ against the ground truth phoneme sequence using the CER  (Character Error Rate) metric.
For the most informative/distinctive representation of the phonemes (to be used in P2V), we collect the samples with the lowest CER. Therefore, we shortlist only those samples where CER is $0$ and randomly pick $\sim 50K$ samples from them, denoted as $D$.

\textbf{Phoneme Vector:} Lastly, for each phoneme $\hat p_{j_i}$ in the predicted phoneme sequence $\hat p$, we trace back to the index range $[l, r]$ in the audio embedding $e$ which corresponded to the prediction of $\hat p_{j_i}$ and define a local vector for this occurrence of $\hat p_{j_i}$ as $\textit{LV} \inparanth {\hat p_{j_i}}=\frac{1}{r-l}\sum_{k=l}^{r} e_k$ (i.e., the average of all the embedding vectors for the audio frames mapped to that phoneme).
We further define a global vector $GV(p) \in \mathbb{R}^d$, $\forall p \in \mcalp$, as the average of all local vectors for $p$ across all samples in the previously defined dataset $D$. We then store $GV(p), \forall p \in \mcalp$ in the P2V Database in Fig.~\ref{fig:model}. 

\begin{figure}[t!]
  \centering
  \includegraphics[width=0.90\linewidth]{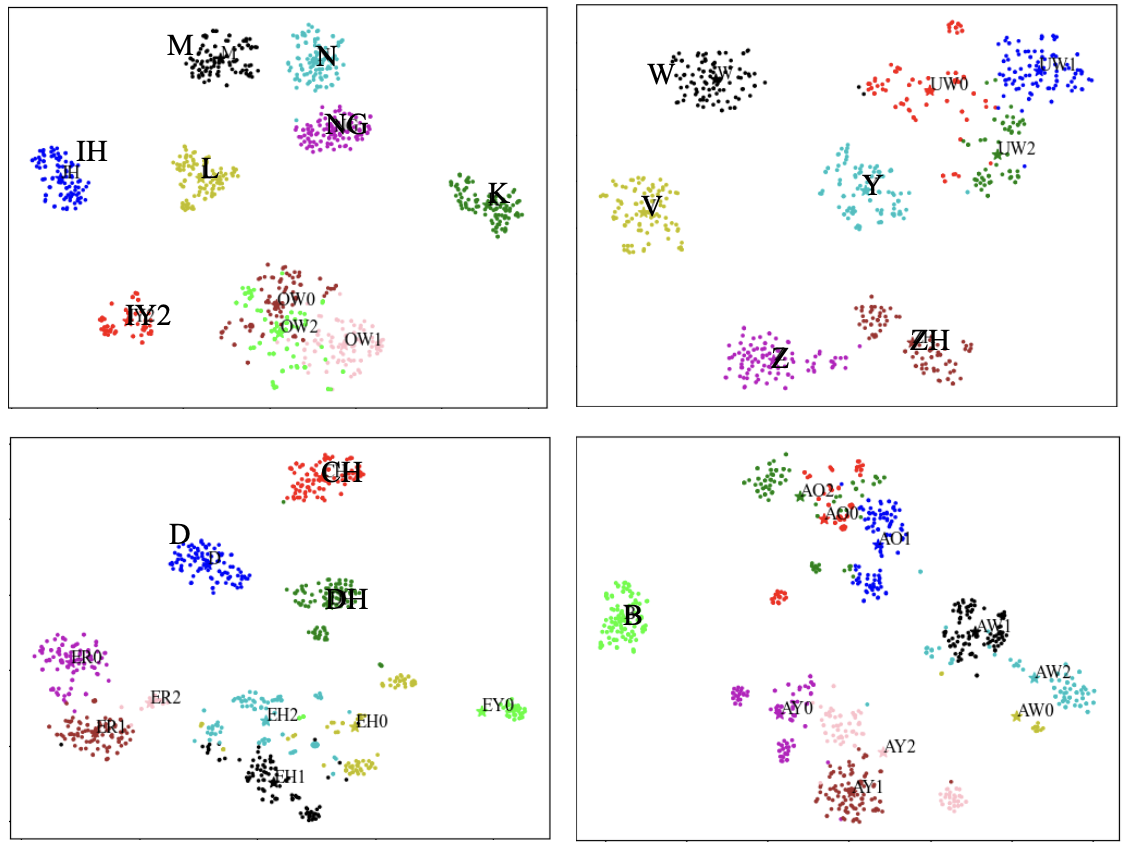}
  \caption{t-SNE visualization of $100$ randomly selected local vectors for various phonemes in different colors: Subset of $10$ phonemes are shown in each subplot.}
  \label{fig:tsne}
\end{figure}

We visualized $100$ randomly selected local vectors $LV(p)$ for different phonemes on the t-SNE plot \cite{JMLR:v9:vandermaaten08a} in Fig. \ref{fig:tsne}. The plot shows how well local vectors are semantically clustered, achieving high intra-class compactness and inter-class separation. At the same time, if we look at the vowel phonemes having the same vowel symbol but with different lexical stress markers such as $\inparanth{OW0,OW1,OW2}$ in top-left subplot of Fig. \ref{fig:tsne}, we find that they have more inter-class closeness compared to other phonemes, but still there is separation with each other. This supports the efficacy of our method in generating effective phoneme vectors for audio-compliant text encoder.

    
\subsection{Verifier}
\label{subsec:Verifier}
We perform the input audio and text matching in the verifier module. 
The module receives the audio embedding $e$ from the audio encoder and the text embedding $f$, obtained by transforming the text phoneme sequence $p$ using P2V.

We generate the cosine similarity matrix of $e$ and $f$ to measure the similarity between audio and text embedding. Since both embeddings come from the same embedding space, we expect a monotonic stepwise alignment pattern for a positive audio and text pair in the cosine matrix, where one phoneme can be associated with one or more consecutive audio frames. We borrow the Dynamic Sequence Partitioning (DSP) algorithm proposed in \cite{nishu2023matching} to obtain this alignment pattern. For further processing, we focus only on the similarity weights along this alignment pattern to enforce the sequential matching of audio and text. Hence, except for the alignment region, we mask other parts of the cosine matrix. We take this masked cosine matrix and perform a dot product with the audio embedding to get the final audio-text agreement matrix of dimension $m\times d$, where $m$ is the phoneme sequence length and $d$ is the embeddding dimension. This output is passed to  a single GRU layer and then to a feed-forward layer which produces a final matching score for the input audio and text pair.


\subsection{Confusable keyword generation}
\label{subsec:confusable_kw_generation}

\begin{figure}[t!]
  \centering
  \includegraphics[width=\linewidth]{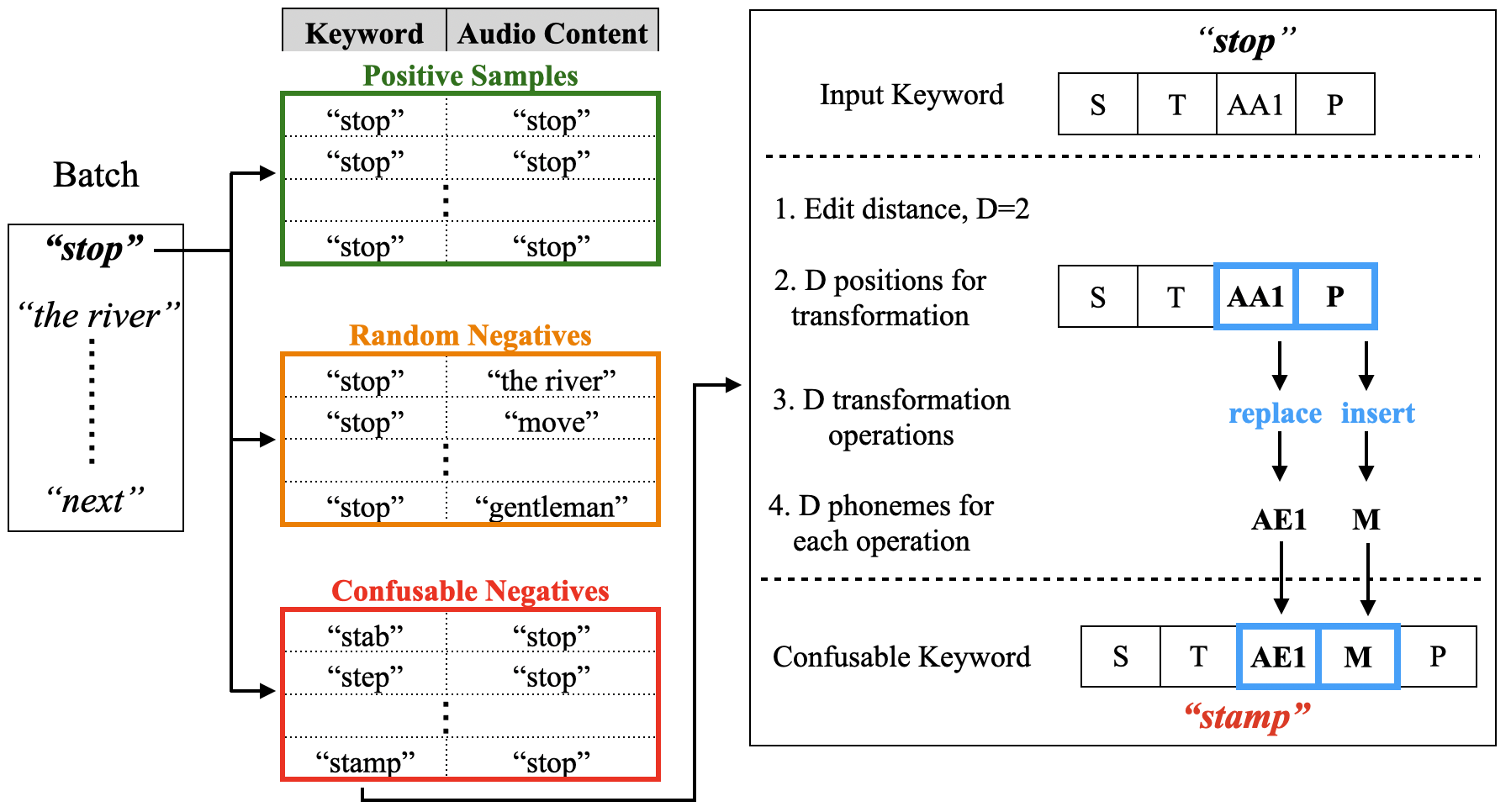}      
 \caption{Overview of the batching scheme for CED model training. A batch is created over keywords and there are $3$ mini-batches for each keyword: positives, random negatives, and confusable negatives. Right block shows the confusable keyword generation steps for a keyword \textit{``stop"}. 
}
 \label{fig:confusable}     
 \vspace{-0.15in}
\end{figure}

False triggers due to phonetic confusability of a user-defined keyword with a similar sounding unintended keyword is a key challenge in the KWS task. Unlike fixed KWS, in flexible KWS, there are no fixed classes for user-defined keywords inside the model and many of such keywords are not even included in the training dataset. Hence, to better handle arbitrary user-defined keywords and empower model with discriminative understanding of phonetic confusability, we design a novel method for auto-generation of confusable keywords as part of the training flow, illustrated in Fig. \ref{fig:confusable}. The generation method is executed in below steps, where input is the keyword and output is a confusable variation of the keyword.
\begin{enumerate}
  \vspace{-0.05in}\item Select an edit distance $\delta$ which denotes the number of phoneme edits in the generated confusable keyword. We suggest using $\delta \in \{1,2,3\}$ in order to generate a hard negative sample. 
  \vspace{-0.05in}\item Randomly select $\delta$ positions in the keyword phoneme sequence, denoted as $u_1, u_2,\dots u_{\delta}$.
  
  \vspace{-0.05in}\item Select $\delta$ transformations as either \textit{replace} or \textit{insert} for each position in $u_1, u_2,\dots u_{\delta}$
  
  \vspace{-0.05in}\item For each position $u_i, i \leq \delta$, randomly select a phoneme which is different from the
  current phonemes at $u_{i}-1$, $u_{i}$, and $u_{i}+1$ in the input keyword. Then apply the transformation at $u_i$ with the selected phoneme.
\end{enumerate}

\section{Experimental Results}
\label{sec:experiments}

\subsection{Datasets}
We  used LibriSpeech \cite{Panayotov2015LibrispeechAA} 
to construct  Libriphrase training and test dataset, following steps from \cite{nishu2023matching, Shin2022LearningAA}. The Libriphrase training dataset is constructed from \textit{train-clean-100/360} and the test dataset from \textit{train-others-500}. Libriphrase test dataset has two parts: Libriphrase Easy (LE) and Libriphrase Hard (LH), detailed in \cite{nishu2023matching, Shin2022LearningAA}. First, we train the audio encoder on longer audio from \textit{train-clean-100/360} and fine-tune on shorter audio from Libriphrase\cite{Shin2022LearningAA}. Then, we train the CED model end-to-end on Libriphrase training dataset. We evaluate the proposed method on both LE and LH. Additionally, we evaluate our method on $10$ short commands from the Google Speech Commands V1 test dataset \cite{Warden2018SpeechCA}.
We  experiment in PyTorch using x86 Linux machines with  NVIDIA V100 GPUs.

\subsection{Training and Evaluation}
 The input audio is processed using 80-channel filterbanks from a 25ms window and a stride of 10ms. The conformer hyper-parameters are \textit{\{6 encoder layers, encoder dimension d=144, convolution kernel of size 3, and 4 attention heads\}}.  We train using Adam optimizer \cite{Kingma2014AdamAM} and transformer learning rate schedule \cite{Vaswani2017AttentionIA} with $5k$ warm-up steps for $150$ epochs. 
 
 

For the end-to-end CED model training, we keep the audio encoder frozen and train the verifier with cross-entropy loss. Our CED model has total $3.8M$ parameters. Our text encoder does not have any additional parameters apart from the G2P model ($0.83M$), as compared to the expensive text encoder (DistilBERT \cite{DistilBERTAD} of $66M$) used in \cite{nishu2023matching}. We employed an exhaustive data batching scheme for CED training, shown in Fig. \ref{fig:confusable}. A training batch of size $32$ is formed over the keywords from the Libriphrase training dataset. And there are three mini-batches (each of size $11$) for each keyword selected in the batch: a positive set, a negative set, and the confusable set where audio samples are same as the positive set but paired with confusable keywords. We evaluate the contribution of confusable keywords by removing them from the training batch and report the results as \textbf{Ours}$^\dagger$ in Table \ref{tab:result}, which shows degradation on both LH and G compared to \textbf{Ours+conf}$^*$.

Evaluation results show that our proposed method outperforms the baselines from \cite{Shin2022LearningAA} and \cite{nishu2023matching} in terms of both, Area Under the ROC Curve (AUC) and Equal-Error-Rate (EER) metric, shown as \textbf{Ours+conf}$^*$ in Table \ref{tab:result}. On the LH dataset, it advances the state-of-the-art results by a significant jump of $10.1 \%$ on the AUC metric and by $38.3\%$ on the EER metric. On the LE dataset, it improves the state-of-the-art baseline results from \cite{nishu2023matching} by $2.05 \%$ on the AUC metric and by $76.9\%$ on the EER metric. Moreover, we measure the generalization of the model on a dataset of different speech characteristics, Speech Commands V1, without any fine-tuning , and compare against baseline \cite{Shin2022LearningAA} which has been evaluated in a similar setup. We find a consistent improvement of $15.9 \%$ on the AUC metric and $50.6\%$ on the EER metric.

\begingroup
    \renewcommand{\arraystretch}{1.1}
        \begin{table}[!t]          
          \centering
          \begin{tabular}{l|l|l|l|l|l|l}
            \hline
            \multirow{2}{*}{Method} & \multicolumn{3}{l|}{\textbf{AUC Score (\%)} $\uparrow$} & \multicolumn{3}{l}{\textbf{EER (\%)} $\downarrow$} \\
            & LH & G & LE & LH & G & LE\\
            \hline
            \cite{Shin2022LearningAA} & 73.58 & 81.06 & 96.7 & 32.9 & 27.25 & 8.42\\
            \cite{nishu2023matching} & 84.21 & - & 97.83 & 23.36 & - & 7.36\\
            \textbf{Ours}$^\dagger$  & 89.2 & 93.16 & \textbf{99.94}  & 18.4 & 14.05 & \textbf{0.8}\\
            
            \textbf{Ours+conf}$^*$ & \textbf{92.7} & \textbf{93.94} & 99.84  & \textbf{14.4} & \textbf{13.45} & 1.7\\

            \textbf{Rel. Imprv}$^{\S}$  & \textbf{10.1} & \textbf{15.9} & \textbf{2.05}  & \textbf{38.3} & \textbf{50.6} & \textbf{76.9} \\
            \hline
          \end{tabular}
          \vspace{0.1 in}
          \caption{Evaluation of CED model on Libriphrase Hard (LH), Libriphrase Easy (LE), and Speech Commands V1 (G) dataset. $(^\dagger)$: our model without confusable module, $(^*)$: our model with confusable module, $(^{\S})$: relative improvement of $(^*)$ from the baseline \cite{nishu2023matching} on LE/LH and \cite{Shin2022LearningAA} on G.}
          \label{tab:result}
        \end{table}
    \endgroup

            


\section{Conclusions}
\label{sec:conclusions}
We have proposed an end-to-end user-defined keyword spotting method based on homogeneous audio-text embedding. We have introduced an audio-compliant text encoder which produces text embedding from the same embedding space as the audio encoder. We also address a key challenge in keyword spotting task, false triggers occurring from the phonetic confusability, by proposing an auto-generation approach for confusable keywords during training. Experimental results show that the proposed method outperforms the state-of-the-art baseline results.

\vfill\pagebreak

\bibliographystyle{IEEEbib}
\bibliography{main}

\end{document}